\pdfoutput=1
\documentclass [10pt,letterpaper]{JHEP3}
\usepackage{amsmath}
\usepackage{bm}         
\usepackage{verbatim}   
\usepackage{graphicx}
\usepackage[all]{xy}
\advance\parskip 1.0pt plus 1.0pt minus 2.0pt   
\def\R{{\mathbb R}}

\def\tr{{\rm tr}}

\def\Z{{\mathbb Z}}

\def\Dslash{{\rlap{\raise 1pt \hbox{$\>/$}}D}}

\title
    {%
    \boldmath
Comments on large-N  volume independence    
 }%

\author
    {
    {
    \def\href#1#2{#2}	
    Erich Poppitz$^1$\footnote{\email{poppitz@physics.utoronto.ca}}~
    and Mithat \"Unsal$^2$\footnote{\email{unsal@slac.stanford.edu}}~
           \\${}^1$Department of Physics, University of Toronto,
    Toronto, ON M5S 1A7, Canada
     \\${}^2$SLAC and Physics Department, Stanford University, Stanford, CA 94025/94305, USA
        }
    }%

\abstract
    {%
   { 
   We study aspects of the  large-$N$ volume independence on $\R^3$$\times$$L^{\Gamma}$, 
   where  $L^{\Gamma}$ is   a  $\Gamma$-site lattice for Yang-Mills theory with adjoint Wilson-fermions.   We find the critical number of     lattice sites above which  the center-symmetry  
   analysis  on $L^{\Gamma}$  agrees with the one on the continuum $S^1$.  
For Wilson parameter  set to one and   $\Gamma$$\geq$$2$,  the two analyses agree. One-loop radiative corrections to Wilson-line 
 masses are finite,  reminiscent of the UV-insensitivity of the Higgs mass  in    deconstruction/Little-Higgs theories. Even for theories with  $\Gamma$$=$$1$,  volume independence in QCD(adj) may be guaranteed to work by tuning one low-energy  effective field theory parameter.  Within the parameter space of the theory, at most three operators of the 3d effective field theory  exhibit 
 one-loop UV-sensitivity.     This opens the analytical prospect to study 4d non-perturbative physics by using 
lower dimensional field theories ($d$$=$$3$, in our example). 
}

      \vspace{5cm}

\hfill published in JHEP01(2010)098
    
    }%

\begin{document}

\section{Volume independence, regularization dependence,  and adjoint fermions}
Consider a  large-$N$ non-abelian asymptotically free gauge theory compactified toroidally on a four-manifold  $M_4$$=$$\R^d$$\times$$(S^1)^{4-d}$.  
 It is well-known since the early 1980s \cite{Eguchi-Kawai, LGY-largeN, BHN} that if center symmetry, or approximate center symmetry  if center is absent, is preserved and if translational invariance 
 is unbroken, to the extent possible in a compactified theory, these theories obey volume independence.\footnote{In the early literature, the necessity of  unbroken translation symmetry is realized  only in  \cite{LGY-largeN}. The possible breaking of translation symmetry is not of only formal interest, see \cite{Bringoltz:2009ym}. If one views   volume independence in the class of  large-$N$ orbifold equivalences  \cite{Kovtun:2007py}, the spontaneous breaking of  discrete translation symmetries in orbifold theories is ubiquitous.} 
  In a lattice regularized theory,  the  reduction to a one-site  model is  known as Eguchi-Kawai (EK) reduction.
     Despite the  
 familiarity with the   necessary and sufficient conditions which validate such an equivalence, it took a long time to 
find a working  example of a four-dimensional gauge theory obeying volume independence in arbitrarily small volumes \cite{Kovtun:2007py}.  Retrospectively, the physical reason is clear.   Until  recently,  for gauge theories formulated on  $M_4$$=$$\R^3$$\times$$S^1$, 
 only thermal compactifications were  considered (with the notable exception of supersymmetric theories). 
 Then, of course, for the class of theories that we have in mind, small-$S^1$ corresponds to a deconfined, center-symmetry broken phase, for which  volume independence is invalid.

 In order to find field theories for  which  volume independence  may work all the way down to arbitrarily small radius, one must therefore  abandon thermal compactification and the use of 
the  partition function   $Z(\beta)$$=$$\tr (e^{-\beta H})$, where $H$ is the Hamiltonian and $\beta$---the inverse temperature.  Instead,  one needs to consider non-thermal circle compactifications, for which fermions are  endowed with translationally  invariant periodic boundary conditions and the twisted partition function 
$Z(L)$$=$$\tr [e^{-L H} (-1)^F]$, 
 where $F$ denotes fermion number and $L$ is the circle size.  
Since thermal and quantum fluctuations are  different 
in a gauge theory,  different choices of boundary conditions for fermions may lead to a  different realization of center symmetry at small radius. With this intuition,  Kovtun, Yaffe and one of us (M.\"U)  showed \cite{Kovtun:2007py} that  a subclass of large-$N$ QCD-like gauge theories satisfies volume independence upon  compactification on $M_4$. Examples of these theories are $SU(N)$ QCD with $N_f < 5.5$ massless adjoint Weyl fermions, for which a fermion-induced stabilization is at work; in what follows we call this class of theories QCD(adj).\footnote{Other examples are SO and Sp theories with adjoint representation fermions in the same range and SO (Sp) theories  with $2 \leq N_f < 5.5$ symmetric (anti-symmetric) representation fermions.}

The analysis of  stability of center symmetry in Ref.~\cite{Kovtun:2007py}   is done in  a 4d continuum.  Let the Wilson line winding around the $S^1$ be
$\Omega(x)$$=$$P e^{i \int A_4dx_4}$.  The 
one-loop potential for its expectation value is:
\begin{equation}
V[\Omega]= (N_f-1)  \frac{2 }{\pi^2 L^3} \sum_{n=1}^{\infty} \frac{1}{n^4} 
|\tr \Omega^n|^2~,
\label{KUY}
 \end{equation}
 and leads to a stable center symmetry for $N_f$$>$$1$.
A similar continuum analysis on  an $S^3 \times S^1$ manifold  also leads to an unbroken center symmetry in any domain where the one-loop analysis is reliable \cite{Unsal:2007fb,Hollowood:2009sy}. In fact, since the first homotopy groups  $\pi_1(S^3$$\times$$S^1)$$=$$\pi_1(\R^3$$\times$$S^1)$$=$$\Z$ coincide, in both cases the effective action or effective one-loop potential is expressed in terms of the Wilson line around the $S^1$ circle, and the result on   $S^3$$\times$$S^1 $ reproduces the $\R^3$$\times$$S^1 $ result by taking an infinite $S^3$ radius.

However, the continuum analysis on $\R^3$$\times$$S^1$ or $M_4$, although suggestive,  does not necessarily imply  that   volume independence will also hold in a lattice regularized  \cite{Bringoltz:2009mi}  or partially lattice regularized theory
 \cite{ Bedaque:2009md}. 
 If we consider replacing $S^1$  with a discrete lattice with $\Gamma$-sites:
\begin{equation}
M_{3, \Gamma}=\R^3 \times L^{\Gamma}  \equiv 
   \R^3 \times \{\Gamma-{\rm sites \;  lattice}\},
\end{equation}
then,  in the continuum limit  
\begin{equation}
\Gamma \rightarrow \infty, \qquad a \rightarrow 0,  \qquad L= \Gamma a= {\rm fixed}~,
\label{cont}
  \end{equation}
 with  $L \Lambda_{\rm QCD} \ll1$ ($\Lambda_{\rm QCD} $ is the strong-coupling scale of the theory), one must reproduce the  continuum  one-loop result  from 
  lattice perturbation theory. Thus, at large $\Gamma$, the two analysis 
   {\it must} agree. Indeed,  as we will show explicitly,  they do so.    
  Ref.~\cite{Bedaque:2009md}  studied the  $\Gamma=1$ theory with the intention of mapping 
  the dynamics of a four-dimensional   gauge theory to a three-dimensional effective field theory. 
  This study has potential importance, since certain tools available for lower dimensional gauge theories can be put at work for the 4d gauge theory by using large-$N$ volume independence.  
 However,   Ref.~\cite{Bedaque:2009md}   reached   somehow puzzling  results, arguing that  
 adjoint fermions do not seem to be sufficient to stabilize the center symmetry of the  theory on 
 $M_{3, \Gamma=1}$.  Ref.~\cite{Bringoltz:2009mi} studied the $\Gamma$-site model 
  both analytically and numerically  by using a lattice-regularized theory and its continuum limit in the non-compact dimensions. It clarified that  the $\Gamma$$=$$1$ model studied in   Ref.~\cite{Bedaque:2009md} 
  was incomplete in the sense of effective field theory, 
   and    for  $\Gamma$$=$$1$ it  concluded that  the  center symmetry was not broken by the order parameter with single-winding Wilson operators (for which  Ref.~\cite{Bedaque:2009md} reported otherwise). This implies that, at least a $\Z_2$ subgroup of $\Z_N$ must be unbroken.  The analysis of Ref.~\cite{Bringoltz:2009mi}  includes the chiral point, i.e., massless fermions, but does  not address the possibility of breaking of the center  to a subgroup  larger than $\Z_2$.   In this work, one of our goals is to fill this gap. 
         A related numerical simulation for massive $SU(3)$  QCD(adj), using staggered fermions, showed that a lattice system which essentially mimics the $\R^3$$\times$$\{\Gamma$$=$$2\}$-site model remains center symmetric for sufficiently light fermions 
 \cite{Cossu:2009sq}.  
 
In this work, prompted by  Refs.~\cite{Bedaque:2009md,Bringoltz:2009mi}, we address  the following questions: 
\begin{itemize}
{\item
How many lattice sites ($\Gamma_{\rm cr}$) along the compact direction are necessary in order for the {\it  center-stability analysis}  on $\R^3$$\times$$L^{\Gamma}  $ to agree with the one on 
 the continuum $\R^3$$\times$$S^1$?   }
 {\item If 
  $\Gamma_{\rm cr}$ is larger than one, is this detrimental for large-$N$  volume independence 
  for volume reduction down to   $\Gamma$$\leq$$\Gamma_{\rm cr}$ ? }
  \end{itemize}

We hope that understanding the answers will eventually lead to developing   analytical  
tools of practical utility to using volume independence of QCD(adj). On the numerical side, there is already important  progress:
 Recent work by Bringoltz and Sharpe \cite{Bringoltz:2009kb} studied   
QCD(adj) on a single-site  lattice, and showed that center symmetry remains  intact in all channels, including the cross-correlation of Wilson loops in different directions.\footnote{It is indeed necessary to check such cross-correlations. The validity of  volume independence requires that the entire center-symmetry group  be unbroken. For example,  the Quenched-EK  model \cite{BHN}   suffers from an exotic breaking of  center symmetry, 
$(\Z_N)^4$$\rightarrow$$(\Z_N)_{\rm Diag.}$ \cite{Bringoltz:2008av}. Thus, even  though  expectation values of Wilson loops $\langle \tr U_{\mu}^{n_{\mu}} \rangle$$=$$0$, 
  loops in different directions are locked so that  the center breaks down to diagonal, for example, $\langle \tr U_{1} U_{2}^{\dagger} \rangle$$\neq$$0$, invalidating the equivalence between the infinite volume theory and the matrix model.   Such exotic breaking is absent in QCD(adj). 
 }
This  provides  numerical evidence  for the equivalence of 
 the single-site matrix model and infinite-volume field theory   in the 
$N$$=$$\infty$ limit.\footnote{For proposals 
which implement  
 EK-reduction by using non-compact matrix models, see for example  
 \cite{Hanada:2009kz, Ishii:2008ib}.  
 For a study which takes advantage of the confined phases of YM theory, 
 see Ref.~\cite{KNN}. 
 }
 
  We  also would like to {\it a priori} state that the results of our analysis for  $\Gamma_{\rm cr}$ are 
 not new and  are  recently obtained in  Ref.~\cite{Bringoltz:2009mi} by using  different methods. 
However, our one-loop potentials  (\ref{G-site3}, \ref{G-site3adj}, \ref{G-site3doubler}) on $M_{3, \Gamma}$ are   new and are particularly useful  to analyze center stability and to check limits, such as the continuum limit,  which must agree with  \cite{Kovtun:2007py}, and the supersymmetric limit 
for the $N_f$$=$$1$ theory, for which it must vanish. Restricted to  $\Gamma$$=$$1$,   these potentials coincide with the chiral limit of the expression given in  Ref.~\cite{Bringoltz:2009}.

\section{Pure Yang-Mills theory  on $ {\bf M_{3, \Gamma}= \R^3 \times L^{\Gamma}}$}
\label{pure}

Consider a  gauge theory with  $[U(N)]^\Gamma$ product gauge-group structure\footnote{The reason we prefer  to use $U(N)$ notation is that it simplifies obtaining the spectrum. Note that the mass spectrum has a natural geometric interpretation in terms  lengths of strings stretched between  appropriately positioned $D$-branes. At large $N$, the $U(1)$ factors are inessential.}
 on 
$M_{3, \Gamma}$$=$$\R^3$$\times$$L^{\Gamma}$, with action: 
\begin{equation}
S=  \sum_{I=1}^{\Gamma}  \frac{L}{g^2} \int  d^3x  \left(\frac{1}{4} \tr |F_{\mu \nu, I}|^2 + \frac{1}{2 a^2} \tr |D_\mu U_I|^2 + \ldots \right)
\label{YM}
\end{equation}
where $F_{\mu \nu}$$\equiv$$||(F_{\mu \nu})^i_{\ j}||$ is the three-dimensional field strength (one for each of the $\Gamma$ $U(N)$ factors). $U_I(x)$ denote the group-valued link fields  between sites $(I, I+1)$. The unitary link field  $U_I(x)$ is local in  $\R^3$, transforms in the $(\Box_I, \overline \Box_{I+1})$ representation of  $[U(N)]^\Gamma$, and can be represented as $U^i_{I \; j} \equiv (e^{i \Phi_I})^i_j$, with $\Phi_I$---an arbitrary real    $N$$\times$$N$  matrix. The trace in (\ref{YM}) is over the indices $i,j=1...N$. The overall normalization of the coupling in (\ref{YM}) will not be essential in our one-loop analysis. 

The product-group theory (\ref{YM})   may be viewed in several different ways, for example, as a non-linear sigma model 
in 3d with a product gauge-group structure (a.k.a.~``theory space," ``moose,"  or ``quiver"). It may also be viewed as a  latticization of the compact circle in    $M_4$$=$$\R^3$$\times$$S^1$. 
As a three dimensional field theory, (\ref{YM}) is non-renormalizable. Thus, the ellipsis are also meaningful in the above formula, and stand for counter-terms which make  the Lagrangian complete as an effective field theory (EFT). Indeed, we will see some counter-terms (and hence EFT parameters) that need  to be added to the action. 

  Under a gauge rotation, the link fields transform as bifundamentals, 
  $U_I(x)\rightarrow g_I(x)U_I(x) g_{I+1}^{\dagger}(x)$, $I$$=$$1 \ldots \Gamma $, and $\Gamma+1$$\equiv$$1$. The  covariant derivative in (\ref{YM}) is 
  $D_\mu U_I$$=$$\partial_{\mu}  U_I +i A_{\mu, I}   U_I - i   U_I  A_{\mu, I+1} $.
   One can build a non-local lattice Wilson line:
  \begin{equation}
  \Omega(x)= U_1 U_2 \ldots U_\Gamma~,
  \end{equation}
 transforming as an adjoint scalar from 3d-point of view:
    $\Omega(x) \rightarrow g_1(x) \Omega(x)g_{1}^{\dagger}(x)$. Thus 
$\tr \Omega^n$ is gauge invariant for any integer $n$.   In what follows, we will calculate the one-loop potential for the gauge invariant Wilson line. Our calculation follows closely 
the  calculation   on  $M_{4, \Gamma}$$=$$\R^4$$\times$$L^{\Gamma}$ by Georgi and Pais \cite{Georgi:1974au} and  on  $M_{d, \Gamma}$$=$$\R^d$$\times$$L^{\Gamma} $  for arbitrary-$d$ by Neuberger \cite{Neuberger}.  

Consider a gauge covariant  background holonomy:
\begin{equation}
\Omega= {\rm Diag } \left( e^{i v_1}, \ldots,   e^{i v_N} \right) ~.
\label{hol}
\end{equation}
  The spectrum of   gauge bosons in the background (\ref{hol}) can be calculated by studying the ``hopping" terms in the Lagrangian (\ref{YM}). 
Although $\tr \Omega^n$  are gauge independent, and we will derive a one-loop potential in terms 
of these operators, the hopping terms in the compact direction are written in terms of 
$U_I$, which are gauge dependent. We choose a gauge in which $\langle U_I \rangle$$=$$U$$=$$\Omega^{1/\Gamma}$, a democratic distribution of holonomy to each link field:
\begin{equation}
\langle U_I \rangle \equiv U =  {\rm Diag } \left( e^{i v_1/\Gamma}, \ldots,   e^{i v_N/\Gamma} \right) ~.
\label{hol2}
\end{equation}
The evaluation of the spectrum of the gauge bosons in this  background follows from expanding the hopping terms in the Lagrangian:
\begin{eqnarray}
L \supset \sum_{I=1}^{\Gamma}  \tr|A_{\mu, I} U_I -U_IA_{\mu, I+1}|^2=  \sum_{I=1}^{\Gamma}  \tr|A_{\mu, I}  -UA_{\mu, I+1}U^{\dagger}|^2~, 
\end{eqnarray}
where in the second formula, we used our $\langle U_I \rangle$$=$$U$ gauge choice. Fourier transforming the  real-space lattice to the momentum Brillouin-zone variable, gives the spectrum of 
gauge bosons:
\begin{equation}
M_{ij}^2(k)= \left[\frac{2}{a} \sin \left(\frac{2 \pi k  +  v_{ij} }{2\Gamma} \right) \right]^2,  ~ v_{ij} \equiv v_i - v_j, ~ k=1, \ldots \Gamma, \; \;  i, j=1, \ldots N~.
\label{spectrum}
\end{equation}
This result is rather intuitive: Setting $v_{ij}=0$ gives the spectrum of bosons in lattice gauge theory; $M(k)= | \frac{2}{a} \sin \frac{ \pi k}{\Gamma} |$, also familiar from deconstruction \cite{Cheng:2001vd, ArkaniHamed:2001nc}. 
 Setting $\Gamma=1$  gives the distance between the eigenvalues of the Wilson line; 
$M_{ij}=\frac{1}{a} |e^{iv_i}-  e^{iv_j}|= |\frac{2}{a} \sin  \frac{v_{ij} }{2}|$, the ``$W$-boson" spectrum, familiar from geometric D-brane pictures; see, e.g.,~\cite{Giedt:2003xr}.

We evaluate the  one loop potential by a hard cutoff scheme in $d=3$; recall that the fourth dimension is already latticized.  The result is: 
\begin{eqnarray}
&&V[ \Omega] = \sum_{k=1}^{\Gamma} \sum_{i,j=1}^N I(v_{ij}) \;,\cr
&&I(v_{ij})= \int_{\Lambda} 
\frac{d^3p}{(2 \pi)^3} \ln \left[ p^2 +  M_{ij}^2(k) \right]=   \frac{ \Lambda |M_{ij} (k) |^2 }{2 \pi^2} -  \frac{ |M_{ij}(k)|^3 }{6 \pi} + \Lambda^3 \times \{M {\rm -independent}\}\;.
\label{pot}
\end{eqnarray}
Zeta-function or dimensional regularization would set the  divergent terms ($\sim$$\Lambda, \Lambda^3$) to zero, but produce identical results for the cubic term (the linearly divergent term in (\ref{pot}) can be recovered as a  $d$$=$$2$ pole in the dimensionally regulated expression). Setting the cubic divergence to zero is not a problem  as we want to know the holonomy dependence of the one-loop potential. However, setting the linear term
to zero is {\it dangerous}, in the sense we will describe below.

There are some important aspects  of the potential (\ref{pot}). 
Although there is a linear divergence for $\Gamma$$=$$1$, and although each term in the $k$-sum for $\Gamma$$\geq$$2$ has divergent holonomy-dependent contributions,  
 it is straightforward to show that, for $\Gamma \geq 2$, $ \sum_{k=1}^{\Gamma} M_{ij}^2 (k)$ is 
 a linearly-divergent constant, independent of $v_{ij}$ \cite{ArkaniHamed:2001nc}. This is, of course, the mechanism that  the deconstruction and Little-Higgs theories use to solve the (little) hierarchy problem.  
 Unlike supersymmetric theories, where loop corrections to the potential from particles of different spins cancel, here, the UV-sensitivity from particles of the same spin  cancels, due to non-locality of the 
 ``Higgs-scalar" in theory space $L^\Gamma$. As in  Ref.~\cite{ArkaniHamed:2001nc}, the one-loop divergent renormalization is to the cosmological constant,  and the Higgs mass is UV-insensitive.
   
  Since the potential (\ref{pot}) is finite for $\Gamma$$\geq$$2$, we rewrite it in a form  more suitable to studying center-symmetry realizations:
  \begin{eqnarray}
\label{G-site}
  V[\Omega, \Gamma] =\sum_{n=1}^{\infty}  V_n(\Gamma, d)
 \; \;  |\tr \Omega^n|^2, \qquad  
 \end{eqnarray}
 where $ V_n(\Gamma, d)$ is the mass for the Wilson line $\tr \Omega^n$. 
For convenience and to make the discussion more general, we give the general 
formula for $ V_n(\Gamma, d)$ on $M_{d, \Gamma}=\R^d \times L^{\Gamma} $:
 \begin{equation}
\label{fourierv}
  V_n(\Gamma, d)   = (- 1)  \frac{c(d)}{a^{d}}  \left\{ \begin{array}{ll}
   \frac{1}{n  (n^2 \Gamma^2 - 4) (n^2 \Gamma^2 - 1) }   & \qquad 
   \R^4 \times L^{\Gamma}  \\
  \frac{\Gamma}{ (n^2 \Gamma^2 - \frac{9}{4}) (n^2 \Gamma^2 - \frac{1}{4}) } & \qquad  \R^3 \times L^{\Gamma}  \\
     \frac{1}{n     (n^2 \Gamma^2 - 1) }  & \qquad \R^2 \times L^{\Gamma} \\
       \frac{\Gamma}{  (n^2 \Gamma^2 - \frac{1}{4}) } &  \qquad  \R^1 \times L^{\Gamma}  
  \end{array}
  \right.~,
 \end{equation}
 where $c(d)$ are  (unimportant) positive prefactors  depending on dimensionality. Eqn.~(\ref{fourierv}) generalizes the $d$$=$$4$ results of  \cite{Georgi:1974au} (given on the top line) and can be obtained by Fourier transforming the potential (\ref{pot}) with respect to $v_{ij}$ (see, e.g., the $\Gamma$$=$$1$ calculation after Eqn.~(\ref{G-site2})).   It is clear that the Fourier coefficients $V_n(\Gamma, d) $ have poles at:
\begin{equation}
   n \Gamma  =\frac{d}{2} , \qquad  n \Gamma  =\frac{d-2}{2} ,
   \label{poles}
\end{equation}
These poles  are an indication that center-symmetry realizations may 
have UV-sensitivity for low $\Gamma$; we note that  the same result was also obtained in  \cite{Bringoltz:2009mi} by a different method.
We now consider several $d$$=$$3$ cases, starting from the continuum $\Gamma$$=$$\infty$, a multi-site lattice with $\Gamma$$>$$3/2$,  and the one-site $\Gamma$$=$$1$ case.

{\flushleft{\bf Continuum on $\R^3$$\times$$S^1$:}}
The continuum one-loop potential  for pure YM theory on 
$\R^3 \times S^1$ is:
 \begin{equation}
V[\Omega]= -  \frac{2 }{\pi^2 L^3} \sum_{n=1}^{\infty} \frac{1}{n^4} 
|\tr \Omega^n|^2~.
\label{cont2}
 \end{equation}
 Note that  the ``masses" for all Wilson lines are negative, 
 \begin{equation}
 V_n <0,  \qquad  1  \leq n  < \infty, \qquad   \Longrightarrow \qquad  \Z_N \rightarrow \Z_1 ~,
 \end{equation}
 which implies that the $\Z_N$ center symmetry  with action $\Z_N$:  $\tr \Omega$$\rightarrow$$e^{\frac{2 \pi i }{N}} \tr \Omega $ is spontaneously broken to the trivial group $\Z_1$.

 {\flushleft{\bf  $\Gamma$$>$$\frac{3}{2}$-site models:} }The one-loop potential for the   $\Gamma$-site model is given by:
\begin{eqnarray}
\label{multisite}
V[\Omega, \Gamma]  = \sum_{k=1}^{\Gamma} \sum_{i,j=1}^N 
- \frac{1}{6 \pi} 
\left|\frac{2}{a} \sin \left(\frac{2 \pi k  +  v_{ij} }{2\Gamma} \right) \right|^3   =  -  \frac{2 }{\pi^2 a^3} \sum_{n=1}^{\infty} \frac{\Gamma}{ (n^2 \Gamma^2 - \frac{9}{4}) (n^2 \Gamma^2 - \frac{1}{4}) } |\tr \Omega^n|^2 \label{G-site3}~.
 \end{eqnarray}
 That it has the correct continuum limit follows from comparing its infinite-$\Gamma$, $L$$=$$a \Gamma$-fixed, limit  to (\ref{cont2}).
For $\Gamma$$>$$\frac{3}{2}$,  the masses of all Wilson lines are negative and this implies the same breaking pattern as in the continuum: 
 \begin{equation}
 V_n (\Gamma>  \frac{3}{2}) <0,  \qquad    1  \leq n  < \infty, \qquad   \Longrightarrow \qquad  \Z_N \rightarrow \Z_1 ~.
 \end{equation}
This shows that for the pure-YM theory (\ref{YM}),  the center-symmetry realization has no one-loop UV-sensitivity for $\Gamma$$>$$\frac{3}{2}$. Thus, the non-renormalizable theory defined by 
 (\ref{YM}) is one-loop complete, in the sense of EFT. 
 
As already noted, the continuum limit  is continuously connected to this multiple-site domain, and the 
 result (\ref{G-site3})  reduces to  (\ref{cont2}) in $\Gamma$$\rightarrow$$\infty$ limit.   Moreover, even for fixed $\Gamma$, if one takes  a large winding number, $n$$\gg$$1$, 
 in   the lattice dispersion relations, only low-lying modes which possess a quasi-continuum dispersion relation, will dominate in their effect. Thus, also in this regime, we expect the continuum behavior to be produced and   indeed, it is.  The same  effect is discussed for massive QCD(adj) in Ref.~\cite{Bringoltz:2009}.

 {\flushleft{\bf  $\Gamma$$=$$1$-site model:}}
 The analytic continuation of Eqn.~(\ref{G-site3}) to the 
$\Gamma$$<$$\frac{3}{2}$ domain is equivalent to the result of the  zeta-function  and dimensional regularization schemes.   This can be also seen more directly by starting with  the potential (\ref{pot})
 in these regularization schemes,  given by:
\begin{eqnarray}
V[\Omega, \Gamma=1]&& =  -  \frac{4}{3 \pi a^3} \sum_{i, j=1}^{N}  
\left |\sin \frac{ v_{ij} }{2} \right|^3   ~.
\label{G-site2}
\end{eqnarray}
Now, we rewrite (\ref{G-site2}), using the  Fourier expansion:
\begin{equation}
\left |\sin \frac{ x}{2} \right|^3 =  \frac{4}{3\pi} +  \frac{3}{2\pi}  \sum_{n=1}^{\infty} \frac{1}{ (n^2 - \frac{9}{4}) (n^2  - \frac{1}{4}) } \cos n x \; ,
\label{identity}
\end{equation}
and dropping holonomy-independent terms, in a way more useful   for analyzing the center symmetry: 
\begin{eqnarray}
\label{onesite}
V[\Omega, \Gamma=1]  =  - {2 \over \pi^2 a^3} \sum_{n=1}^\infty  \frac{1}{ (n^2 - \frac{9}{4}) (n^2  - \frac{1}{4}) } \sum\limits_{i,j=1}^N \cos  {n v_{ij}}   
 =    \sum_{n=1}^{\infty}  V_n(\Gamma=1, 3)
 \; \;  |\tr \Omega^n|^2. \qquad  
 \end{eqnarray}
As already mentioned, both  these regularization schemes are dangerous as they set terms 
dependent on the Wilson line to zero.  This is an indication that for  $\Gamma$$=$$1$, the non-renormalizable theory defined by  (\ref{YM}) is not a complete theory in the sense of EFT, even at the one-loop order of the calculation.

With this warning, let us continue our examination of the symmetry realization of the $\Gamma$$=$$1$ theory, by using  (\ref{onesite}). Then, we would deduce:
\begin{equation}
V_1(\Gamma=1) >0, \; \;   V_{n \ge 2} (\Gamma=1) <0  \qquad   \Longrightarrow \qquad  \Z_N \rightarrow \Z_2 ~,
\label{deceptive1}
 \end{equation}
 which implies:
\begin{equation}
\Big\langle \frac{1}{N} \tr \Omega^{2n+1} \Big\rangle=0, \qquad \Big\langle \frac{1}{N} \tr \Omega^{2n} \Big\rangle \neq 0.
\end{equation}
This conclusion is, of course, strange. We expect     center 
symmetry to be broken completely, not only partially, in the deconfined phase of pure-YM theory.  Clearly, this must be an artifact of $\Gamma$$=$$1$. But why? 

This question is already answered by Bringoltz \cite{Bringoltz:2009mi}. Here, we follow his arguments. For the $\Gamma$$=$$1$ site theory, the operator with $n$$=$$1$ winding number, $|\tr \Omega|^2$ has a one-loop UV-sensitivity---its mass is linearly UV-divergent.   On the other hand, for the rest of the operators
 $|\tr \Omega^n|^2$ which satisfy  $\Gamma n$$>$$\frac{3}{2}$,  one does not expect any one-loop divergences. This means that the $\Gamma$$=$$1$ theory viewed as an effective field theory (EFT) is one-loop incomplete. One needs to add   a counter-term:  
\begin{equation}
\delta L_{c.t.}= ( c_1 \Lambda + b_1) |\tr \Omega|^2 \;, 
\end{equation}
with  coefficient $c_1$  chosen to absorb the linear divergence and   $b_1$---a low-energy parameter. 
By a choice of $b_1$, we can make the  sign of the pre-factor of the  $|\tr \Omega|^2$ operator the  same as in the continuum or $\Gamma$$>$$\frac{3}{2}$ theories, ensuring that:
\begin{equation}
[b_1 +V_1(\Gamma=1) ]<0, \; \;   V_n (\Gamma=1) <0,  \qquad    2  \leq n  < \infty, \qquad   \Longrightarrow \qquad  \Z_N \rightarrow \Z_1 
 \end{equation}
so that  the unbroken $\Z_2$    symmetry reduces  to $\Z_1$.

To summarize, the realization of the center symmetry for the $\Gamma$$=$$1$ theory is  UV sensitive, i.e. depends on the regularization used. At one loop, this is reflected in the fact  that whether the center is broken down  to  $\Z_2$   or  $\Z_1$ depends on the sign of  $b_1 +V_1$, where $b_1$ is a free parameter of the one-site theory. The continuum result is only reproduced for some values of $b_1$, reflecting this UV sensitivity.

\section{QCD(adj) with Wilson fermions 
on $ {\bf M_{3, \Gamma}= \R^3 \times L^\Gamma}$}
{\bf Continuum $ \R^3 \times S^1$:}
In the continuum,  the adjoint fermion endowed with periodic boundary conditions stabilizes the center symmetry.  As   (\ref{KUY}) shows, for $N_f$$>$$1$ 
 the masses of all   Wilson lines are positive and this is the reason that   the $\Z_N$ center symmetry is preserved at arbitrarily small $S^1$:
 \begin{equation}
 V_n >0 ,  \qquad  1  \leq n  < \infty, \qquad   \Longrightarrow \qquad  \Z_N   \; 
 {\rm intact }~.
 \end{equation}
 
 In this Section,  we would like to study the effect of fermions for the theory on 
$M_{3, \Gamma}$. Following \cite{Bedaque:2009md, Bringoltz:2009mi}, we use 
Wilson fermions.  The spectrum of the 
 adjoint fermions  in the background holonomy (\ref{hol}, \ref{hol2}), using the same notation as for the bosons, is given by: 
\begin{equation}
M_{ij}^2 (k) =\left[  m + \frac{r}{a} 2 \sin^2\left( \frac{ 2 \pi k +  v_{ij} }{2\Gamma}  \right) 
\right]^2 + \frac{1}{a^2} \sin^2 \left( \frac{ 2 \pi k +  v_{ij} }{\Gamma}  \right)  , ~ k=1, \ldots \Gamma, \; \; i,j=1, \ldots N 
\label{specfer}
\end{equation}
Here, $m$ is the bare fermion mass, $r$ is the Wilson parameter to lift the doublers present when $r$$=$$0$ (set, for example, $m$$=$$0$, $v_{ij}$$=$$0$ for convenience; then, apart 
from $k$$=$$0$, modes near $k$$=$$\frac{\Gamma}{2}$ also yield a gapless fermionic excitation). In general,
for each discretized dimension, each corner of the Brillouin zone contributes an extra species. 
This, of course, can be circumvented by the  (chiral symmetry violating) Wilson term,  
which guarantees that these states become infinitely heavy  and do  not contribute  in the continuum limit.\footnote{The price one pays is having to tune the bare mass to cancel linearly divergent contributions to the fermion mass. However, this issue will not be important 
for our one-loop consideration.} The holonomy-dependent part of the one-loop potential is:
\begin{eqnarray}
&&V[ \Omega] = -N_f \sum_{k=1}^{\Gamma} \sum_{i,j=1}^N \left(  \frac{ M_{ij}^2 (k) \Lambda }{2 \pi^2} -  \frac{ |M_{ij}(k)|^3 }{6 \pi} \right).
\label{potfer}
\end{eqnarray}
There are two limits of this expression which are particularly easy to study, and since we want to demonstrate a matter of principle regarding the UV-sensitivity of the  realization of center symmetry, we will restrict our attention to these simple cases (however, these cases  capture   general aspects of the answer). 

{\flushleft{\bf Case I:}}
First, take $m$$=$$0$ and $r$$=$$1$. The spectrum reduces to: \begin{equation}
M_{ij}^2(k)= \left[\frac{2}{a} \sin \left(\frac{2 \pi k  +  v_{ij} }{2\Gamma} \right) \right]^2\; .
\label{spectrum2}
\end{equation}
This spectrum coincides with the one for bosons, similar to the continuum analysis, and, 
because of  the choice of $r$=$1$, it is  doubler-free.  This property of one-dimensional 
Wilson fermions was also observed in the context of deconstruction of supersymmetric theories 
\cite{Giedt:2003xr} and is particularly useful in checking the cancellation of loop contributions of gauge boson and one-fermion species.

With this choice of parameters, 
the one-loop potential for the   $\Gamma$-site  QCD(adj) is:  
\begin{eqnarray}
\label{G-site3adj}
V[\Omega, \Gamma]&& =  (N_f-1) \frac{2 }{\pi^2 a^3} \sum_{n=1}^{\infty} \frac{\Gamma}{ (n^2 \Gamma^2 - \frac{9}{4}) (n^2 \Gamma^2 - \frac{1}{4}) } |\tr \Omega^n|^2~.
 \end{eqnarray}
For $\Gamma$$>$$\frac{3}{2}$,  the masses for all Wilson lines are positive 
 and this implies an unbroken center symmetry at $N_f$$>$$1$, exactly as in the continuum: 
  \begin{equation}
 V_{n \ge 1} (\Gamma>  \frac{3}{2}) >0 \qquad   \Longrightarrow \qquad  
 \qquad  \Z_N   \; 
 {\rm intact }~.
 \end{equation}
 Thus, with this choice of parameters, there is no UV-sensitivity of the center symmetry realization, and the 
 one-loop potential on  $ \R^3 \times L^{\Gamma}$ coincides with the one in continuum  $ \R^3 \times S^1$. 
 
However, for $\Gamma <\frac{3}{2}$,  as it was the case with the pure Yang-Mills theory (flipped in overall sign with respect to  (\ref{deceptive1})), we have:
\begin{equation}
V_1(\Gamma=1) <0, \; \;   V_{n \ge 2} (\Gamma=1) >0 \qquad   \Longrightarrow \qquad  \Z_N \rightarrow \Z_1~.
\label{deceptive2}
 \end{equation}
The reason for the complete breaking   $\Z_N \rightarrow \Z_1$  is that the $V_1$ term 
 dominates the sum.  A sufficiently negative $V_1$ is capable of breaking the center completely even if $V_{n \geq 2} >0$.\footnote{The converse statement is not true. If   $V_{n \geq 2}$ are negative, regardless of the value of $V_1>0$, only a $\Z_2$ subgroup of the center can be restored, not 
 $\Z_N$. In such a case, the  full center symmetry can only be  restored with the tower of  the double-trace deformation of Ref.~\cite{Unsal:2008ch}.}
 However, this can be cured by one counter-term and one judicious choice of the EFT parameter, 
 of the form:
\begin{equation}
\delta L_{c.t.}= (c_1 \Lambda+ b_1) |\tr \Omega|^2 ~.
\end{equation}
{\bf As in the pure gauge theory, see discussion in the end of Section  \ref{pure},} the choice of the EFT parameter, which makes the result on $\Gamma=1$-site model agree with the continuum, is the 
one for which $b_1$$+$$V_1(\Gamma=1)$$>$$0$ is positive. With this choice, we  have:
\begin{equation}
[b_1 +V_1(\Gamma=1) ]>0, \; \;   V_{n \ge 2} (\Gamma=1) >0 \qquad   \Longrightarrow \qquad  \Z_N   \; 
 {\rm intact }~.
 \end{equation}

 {\flushleft{\bf Case II:}} Take $m$$=$$0$ and $r$$=$$0$.  Since $r$$=$$0$ and only one dimension is latticized,  there is only one doubler and it is not lifted. The spectrum is:
  \begin{equation}
M_{ij}^2 (k) =
\frac{1}{a^2} \sin^2 \left( \frac{ 2 \pi k +  v_{ij} }{\Gamma}  \right)  , \qquad k=1, \ldots \Gamma, \; \; i,j=1, \ldots N ~.
\end{equation}
The gauge-boson- and fermion- induced one-loop potential is:
\begin{eqnarray} \label{G-site3doubler}
V[\Omega, \Gamma]  =  \frac{2 }{\pi^2 a^3} \sum_{n=1}^{\infty}  \left(  \frac{ \Gamma N_d N_f }{ (n^2 \Gamma^2 -  9) (n^2 \Gamma^2 - 1) }-    \frac{ \Gamma }{ (n^2 \Gamma^2 -  \frac{9}{4}) (n^2 \Gamma^2 - \frac{1}{4})  }  \right) |\tr \Omega^n|^2,  
\end{eqnarray}
where $N_d$$=$$2$, and $N_f$ is the number of the adjoint Majorana fermions. Generalizing this analysis to $\R^d \times L^{\Gamma}$, 
we observe that the   fermionic contribution contains terms such as:
\begin{equation}
V_n(\Gamma) \supset \frac{1}{(n^2 \Gamma^2 -d^2)(n^2 \Gamma^2 -(d-2)^2)}~,
\end{equation}
for $d>2$, and  has poles at:
\begin{equation}
n\Gamma=d, \qquad n\Gamma=d-2. 
\end{equation}
The  result for $d$$=$$4$ is also obtained long ago in Ref.~\cite{Georgi:1974au}. 
For $d$$=$$3$ and $\Gamma$$=$$1$, the operators with $n$$\leq$$3$ will acquire divergences. These are 
$|\tr \Omega^n|^2$, $n$$\leq$$3$. 
 Examining the parameter space of the one-loop potential (\ref{potfer}) in detail, this is the maximal amount of fine-tuning needed.   For  $\Gamma$$>$$3$, the one-loop potential is UV-insensitive.  
 
{\flushleft {\bf Remark:}} We expect that on an asymmetric lattice, which mimics 
the gauge theory on  $\R^3$$\times$$\{\Gamma$$=$$1\}$-site and the Wilson parameter set to $r$$=$$1$, the center symmetry will not  be broken.  A way to argue this is the following: The lattice regularization is   more similar to hard cut-off than dimensional or zeta-function regularizations.   
 In the continuum, let us fix the UV cut-off and determine the center symmetry realization with it.  
 Then, we obtain, for QCD(adj) with $N_f$ flavors: 
 \begin{eqnarray}
\label{G-site3adj2}
V[\Omega, \Gamma, \Lambda]&& =  (N_f-1)\left\{   \left( \frac{\Lambda}{  \pi^2 a^2}  -  \frac{32 }{15\pi^2 a^3}   \right) |\tr \Omega|^2
+ \frac{2 }{\pi^2 a^3} \sum_{n=2}^{\infty} \frac{1 }{ (n^2  - \frac{9}{4}) (n^2  - \frac{1}{4}) } |\tr \Omega^n|^2 \right\}.
 \end{eqnarray}
For $\Lambda > \frac{32}{15a}$ and $N_f$$>$$1$, the $\Z_N$ center symmetry is stable, and  for $N_f$$=$$0$, the center symmetry breaks down to    $\Z_1$.     Clearly, the realization of center-symmetry depends on the regulator;  we expect that  the lattice regulator   is the one ensuring  stability of the center in the  $1^4$-lattice theory  for QCD(adj) found in    \cite{Bringoltz:2009kb}.

\section{Prospects and open problems}

We have shown how center-stabilization is achieved on $\R^3$$\times$$L^\Gamma$ in QCD(adj).  For the particular choice of Wilson parameter $r$$=$$1$, the center symmetry 
is preserved for any $\Gamma$$\geq$$2$ and $N_f$$>$$1$.   For  $\Gamma$$=$$1$,  center symmetry is preserved by  a choice of a single low energy parameter, reflecting the already mentioned regulator dependence of the center-symmetry realization on the lattice. 

Thus, for $\Gamma$$=$$1$ UV sensitivity of the potential for the holonomy sets in already at one loop, while the $\Gamma$$>$$1$ theory has no one-loop UV sensitivity. This result is in accord with past studies of deconstrucion of five-dimensional field theories \cite{ArkaniHamed:2001ed}, which noted    that  UV-insensitivity of the mass term in a theory with $\Gamma$ sites holds up to $\Gamma$-loop level and that strict all-orders UV-insensitivity is only recovered in the $\Gamma$$\rightarrow$$\infty$ limit.

Center-symmetry stabilization,  although established 
by a perturbative calculation, has important implications for the full, perturbative and  non-perturbative,
dynamics of the theory. Consider for example, the $\Gamma$$=$$1$ model.  Inspecting the 
mass spectrum (\ref{spectrum})   by using  a center symmetric holonomy with 
$v_i$$=$$\frac{2 \pi}{N} i$, we find $M_{ij}=|\frac{2}{a} \sin  \frac{\pi (i-j)}{N}|$, which is equivalent to the 
spectrum in a lattice gauge theory with an effective  box size $ L_{\rm eff }$$=$$Na$.  
This observation is at the heart of volume independence. It demonstrates how the classical   zero mode of the KK-tower of states along the compact dimension  is capable, {\it at large-$N$}, of generating a  Brillouin zone of its own. The ``zero-mode" matrix mode, quantum-mechanically and to all orders in perturbation theory, generates  states with energies $0, \frac{ 2 \pi}{Na},  \frac{2
(2\pi) }{Na}, \ldots \frac{\pi}{a}$.  Although the high energy modes with energies of order 
$\frac{\pi}{a} \gg \Lambda_{\rm QCD}$ are weakly coupled, the low energy modes with  $\frac{2\pi}{aN} \ll \Lambda_{\rm QCD}$ will necessarily 
be strongly coupled. 
This is how a lower dimensional theory on   $\R^3$$\times$$\{\Gamma$$=$$1\}$-site model  
can reproduce the non-perturbative  dynamics of one-higher dimensional theory. 

This discussion easily  generalizes to a $\Gamma$-site model, and  
on a center symmetric background, the effective size of the space reads
\begin{equation}
L_{\rm eff } =\Gamma a N = LN. 
\end{equation}
 This combination makes it manifest that spatial size $L=\Gamma a$ and 
number of colors $N$ are on the same footing, and taking either to infinity is equivalent to 
{\it decompactification}.   So long as  $\Gamma a N \Lambda \gg 1$, the center symmetric theory will produce the four-dimensional gauge 
dynamics.   The importance of the   $L_{\rm eff }$$=$$L N$ scale was realized in 
\cite{Unsal:2008ch} in determining the semi-classical domain 
and soon after  in Refs.~\cite{Meisinger:2009ne, Hollowood:2009sy} in one-loop potentials of massive QCD(adj). 

 Unlike deconstruction, where 4d dynamics can be produced only in the large $\Gamma$ limit,  in our formalism, even with $\Gamma$$=$$1$, the 4d dynamics can be produced in the large-$N$ limit. However,  in our case, the fields of our lower dimensional theory are independent of the coordinate of the compact  space, and it is a genuinely lower dimensional (3d, in our example) theory, which is equivalent to the 4d theory. This possibility  strictly relies on a quantum mechanical condition, i.e,  unbroken center-symmetry, as opposed to being a classical effect, as in deconstruction. The fact that a 4d-theory is non-perturbatively  
equivalent to a 3d (or even lower dimensional) theory makes us optimistic for the analytical utility of our formalism.

 Apart from aiming to show that center-stabilization and  large-$N$ volume independence work  both in the continuum  \cite{Kovtun:2007py} 
 and even for the $1^4$-lattice theory  \cite{Bringoltz:2009kb} (which is the original Eguchi-Kawai dream), 
  attempts to make analytical use of volume independence may  further our understanding of gauge theories. Some topics of interest are: 
\begin{itemize}
\item{\it Conformal field theories:} Consider a compactified CFT. How does volume independence avoid the only length scale in the problem? What are  the implications of volume independence in the AdS/CFT context? 
\item{\it Perturbation theory:} How can results independent of 
 compactification radius arise within perturbation theory? 
\item{\it Large-N matrix models and quantum mechanics:} The combination of 
Refs.~\cite{Kovtun:2007py,Bringoltz:2009kb} tells us that volume independence works both in continuum and 
lattice, and the reduced theories are equivalent to the corresponding gauge theory on $\R^4$. 
Can  one build new analytical methods to study large-$N$ matrix quantum mechanics, and perhaps, even solve it? 
\end{itemize}

 \acknowledgments 
  We thank Barak Bringoltz and Larry Yaffe  for useful discussions. 
 This work was supported by the U.S.\ Department of Energy Grant DE-AC02-76SF00515 and by the National Science and Engineering Council of Canada (NSERC).

 {\flushleft {\bf Note added:}} A recent preprint  by Bringoltz studies   massive and massless QCD(adj) 
 on $\R^3$$\times$$\{\Gamma$$=$$1\}$ by using lattice regularization, and by taking the continuum limit in the 
 non-compact directions \cite{Bringoltz:2009}. The results of our   $\Gamma$$=$$1$ limit are in agreement with his.

\end{document}